# Explainable Automated Machine Learning for Credit Decisions: Enhancing Human Artificial Intelligence Collaboration in Financial Engineering

*Marc Schmitt* [a, b]

[a] *Department of Computer Science, University of Oxford, UK*
[b] *Department of Computer & Information Sciences, University of Strathclyde, UK*

**Abstract**

This paper explores the integration of Explainable Automated Machine Learning (AutoML) in the realm of financial engineering, specifically focusing on its application in credit decision-making. The rapid evolution of Artificial Intelligence (AI) in finance has necessitated a balance between sophisticated algorithmic decision-making and the need for transparency in these systems. The focus is on how AutoML can streamline the development of robust machine learning models for credit scoring, while Explainable AI (XAI) methods, particularly SHapley Additive exPlanations (SHAP), provide insights into the models' decision-making processes. This study demonstrates how the combination of AutoML and XAI not only enhances the efficiency and accuracy of credit decisions but also fosters trust and collaboration between humans and AI systems. By analyzing real-world data and case studies, this paper illustrates the practical applications and benefits of this integration. The findings underscore the potential of explainable AutoML in improving the transparency and accountability of AI-driven financial decisions, aligning with regulatory requirements and ethical considerations.

## 1  Introduction

Artificial intelligence (AI) and machine learning (ML) are pivotal for the digital economy of the future as they drive innovation and efficiency across various sectors, enabling personalized services, optimized operations, and data-driven decision-making (Gershov et al., 2024; Jaenal et al., 2024; Schmitt, 2020). These technologies are fundamental in transforming industries, fostering economic growth, and creating new market opportunities in an increasingly data-centric world (He et al., 2024; Romero & Ventura, 2010).

Finance is among the sectors most impacted by AI/ML. Notable applications include algorithmic trading, investment decisions, fraud detection, customer service, and risk management (Doumpos et al., 2023; Mita & Takahashi, 2023; Zhang et al., 2024). In particular, AI-driven credit decisions, where ML algorithms assess the creditworthiness of individuals or businesses more accurately and efficiently have been subject to strong interest in academia and industry (FUSTER et al., 2022; Schmitt & Cummins, 2023).

AI/ML significantly benefits P2P lenders, FinTech companies, and traditional banks by enhancing the precision of credit scoring and default probability estimation for applicants (Doumpos et al., 2023; Oreski, 2023; Schmitt, 2020). Accurate predictions of credit risk allow financial institutions to minimize potential credit losses (Stiglitz & Weiss, 1981) and avoid the misclassification that leads to lost revenue from incorrectly assessed low-risk customers. Machine learning is crucial in addressing these challenges, particularly in mitigating the adverse selection problem in loan applicant evaluation (Jordan & Mitchell, 2015).

The digital economy is gradually drifting away from manual executions towards either automation or augmentation (Schmitt, 2023; Shepherd & Majchrzak, 2022; Tschang & Almirall, 2020). Human-AI collaboration in the context of credit risk decision-making represents a significant advancement in the financial sector.

Automated Machine Learning (AutoML) is a core element of human-AI collaboration. AutoML automates the process of applying machine learning to real-world problems. It encompasses a suite of techniques and methodologies that aim to reduce or eliminate the need for skilled data scientists to perform complex tasks such as feature selection, model selection, hyperparameter tuning, and model validation. AutoML systems effectively democratize machine learning by making it more accessible to non-experts. However, it also serves as a tool for faster prototyping and benchmarking. The emergence of AutoML is pivotal in addressing the growing demand for machine learning applications while ensuring consistent and robust model performance (Schmitt, 2023).

By combining the analytical prowess of AI with the nuanced judgment of human experts, this collaborative approach enhances the accuracy and fairness of credit assessments. AI algorithms can process vast amounts of data rapidly, identifying patterns and risks that might escape human analysis. However, human intervention is crucial for interpreting ambiguous cases, considering unique circumstances, and ensuring ethical decision-making. This synergy reduces the likelihood of biases inherent in purely automated systems and allows for more tailored and equitable credit decisions. Furthermore, human-AI collaboration facilitates continuous learning, where human feedback helps in refining AI models, leading to smarter and more reliable credit risk evaluations over time. This partnership not only boosts efficiency but also fosters trust and transparency in financial decisions, benefiting both lenders and borrowers.

The black-box nature of AI/ML models is problematic in fields where regulatory compliance and trust are paramount (Rudin, 2019; Saeed & Omlin, 2023; Schmitt, 2020). Understanding the rationale behind credit score-based decisions is crucial, given their significant consequences for individuals and businesses. Thus, explainable AI (XAI) has become increasingly important, ensuring that the decision-making process is transparent and understandable, both for the experts managing the systems and for the individuals affected by these credit decisions (Schmitt & Cummins, 2023).

The objective of this paper is to explore the integration of AutoML and XAI to enhance human-AI collaboration in the realm of credit decision-making. This investigation aims to demonstrate how the synergy between AutoML and XAI can lead to more informed, equitable, and reliable credit assessments, ultimately fostering a more collaborative and trustful relationship between human decision-makers and AI systems. This paper utilizes SHAP, the foremost XAI method, for its global interpretability in attributing individual feature contributions to specific prediction instances (Chen et al., 2023).

The structure of this article is as follows. Section 2 – methods and materials – builds the foundation of this study by providing the background in AutoML, explainable AI (Shapley Values), and outlining the experimental design. Section 3 – results and analysis – presents the empirical results of this study, including a first analysis. Section 4 discusses several aspects of this study and integrates the findings in the context of explainable AutoML for human-AI collaboration. Implications for practice, limitations, and future research directions will also be presented. Section 5 ends with a conclusion.

## 2  Methods and Materials

### 2.1  AutoML

Automated Machine Learning (AutoML) represents a significant advancement in the field of artificial intelligence, streamlining the process of developing machine learning models. The core functionality of AutoML lies in its ability to automate several critical steps in the machine-learning pipeline. This includes the automatic selection of the most appropriate algorithms and models based on the given data, optimizing hyperparameters to enhance model performance, and conducting feature selection and engineering to improve data quality and relevance. AutoML tools also typically incorporate model validation and evaluation techniques, ensuring the accuracy and reliability of the generated models. By handling these intricate and often technical aspects of model building, AutoML allows for a more efficient and less error-prone development process, making sophisticated machine learning models more attainable and practical for a wider range of applications. According to current research, the H2O AutoML framework stands out as one of the most advanced AutoML solutions available. Recent benchmark studies have demonstrated its exceptional performance in both classification and regression tasks (Gijsbers et al., 2019; Schmitt, 2023; Truong et al., 2019). See Figure 1 for the AutoML setup used in this study.

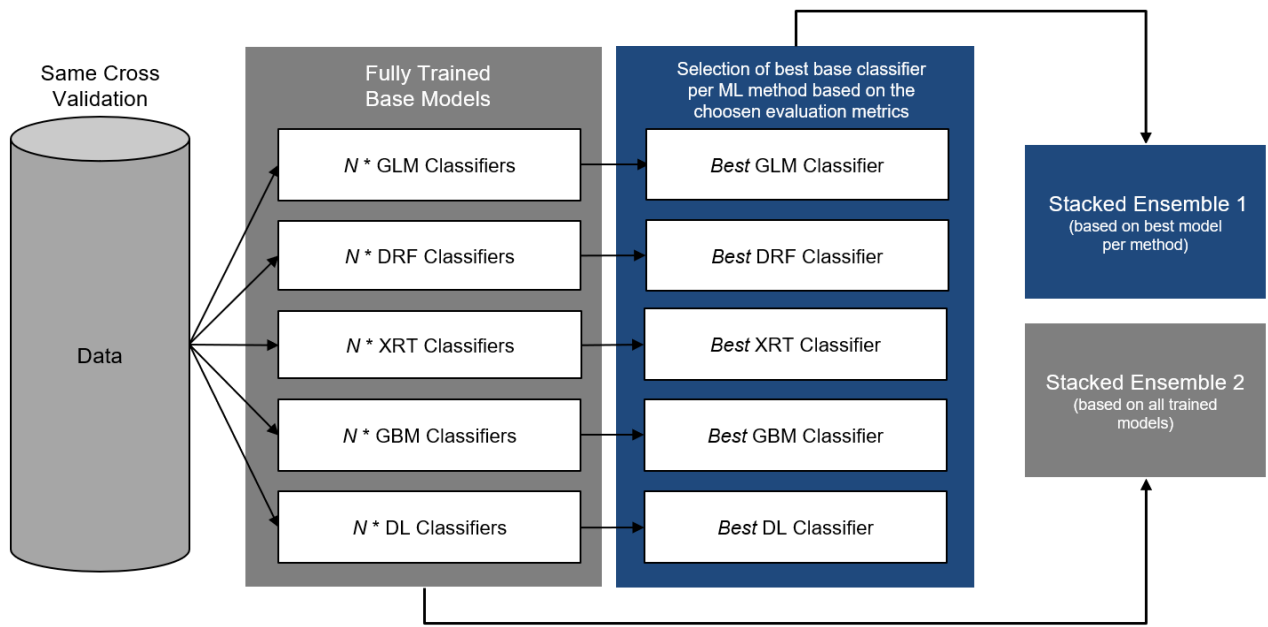

**Figure 1.** The AutoML framework trains several base learners and in a subsequent step combines those into two different stacked ensembles. One stacked ensemble is based on all previously trained classifiers, the other takes only into account the best classifier of each ML family (GLM, DRF, XRT, GBM, DL). AutoML automatically ranks the outcomes based on the chosen evaluation metrics. All final models can be accessed via a "Leaderboard" upon finalization.

The AutoML framework trains a diverse set of base classifiers as well as two stacked ensembles. H2O's stacked ensemble method, a supervised machine learning approach, employs stacking to identify the most effective blend of predictive algorithms. This technique involves a meta-learner, specifically a non-negative GLM, which is trained to ascertain the optimal mix of base learners for enhanced predictive accuracy (Ledell & Poirier, 2020).

Predictive modeling techniques (base classifiers) used in this study:

- **Generalized Linear Model (GLM):** Generalized Linear Model extends linear regression by allowing the response variable to have an error distribution other than the normal distribution. GLM encompasses various statistical models (like logistic regression, and Poisson regression) by linking the expected value of the response variable to the linear predictors via a link function. GLMs are widely used for modeling different types of data and are fundamental in statistical analysis.

- **Distributed Random Forest (DRF):** Distributed Random Forest is an ensemble learning method that constructs multiple decision trees during training, outputting the mode of the classes (classification) or mean prediction (regression) of the individual trees. "Distributed" implies that the computation is parallelized across computing resources, making it efficient for handling large datasets. DRFs are robust to overfitting and are effective in both classification and regression problems.
- **Extremely Randomized Trees (XRT):** Extremely Randomized Trees is an ensemble learning technique like Random Forest. The key difference lies in how the splits are chosen. XRT randomizes the selection of both the features and the thresholds for splitting, leading to more diversified trees. This randomness can sometimes lead to higher model accuracy, particularly in cases with a lot of noise in the data.
- **Gradient Boosting Machine (GBM):** Gradient Boosting Machine is an ensemble learning technique that builds a model in a stage-wise fashion by combining the predictions of several base estimators. It optimizes a differentiable loss function, using the gradient descent algorithm. Each new model incrementally improves upon the previous ones, often using decision trees as the base learners. GBMs are particularly known for their effectiveness in classification and regression tasks.
- **Deep Learning (DL):** Deep Learning is a subset of machine learning that employs neural networks with multiple layers to model complex patterns in data. It is characterized by its ability to learn hierarchical representations of data, utilizing a backpropagation algorithm for training the network. Deep Learning has been highly successful in tasks such as image and speech recognition, natural language processing, and playing complex games.

## 2.2  SHAP Values

SHapley Additive exPlanations (SHAP) is a game theory-based approach for explaining the output of machine learning models and has emerged as a leading XAI method to understand feature contributions (Chen et al., 2023). SHAP values assess the impact of individual features, such as credit amount, employment status, and payment history, on credit decisions. They do this by calculating the average marginal effect of each feature across all possible combinations, assigning a Shapley value that indicates the feature's influence on shifting the model's prediction from a baseline. Thus, SHAP is a crucial tool for financial institutions, enabling them to transparently explain credit decisions made by AI/ML models. However, it's noteworthy that computing SHAP values can be resource-intensive, particularly for high-dimensional datasets, due to the complexity of summing over all feature subsets.

## 2.3 Experimental Design

This study aims to enhance the digital economy by refining credit risk models used by financial institutions. First, by improving the accuracy of these models in assessing client default risk, we can better guide decisions related to loans and credit card approvals. Second, it is vital to integrate Explainable Artificial Intelligence (XAI) for greater transparency and understanding of model decisions. Both accuracy and explainability are essential in aligning credit risk decisions with the dynamic demands of the digital economy. All stages, including data preparation, preprocessing, model fitting, and evaluation, were conducted using RStudio, a popular Integrated Development Environment (IDE) for data science and machine learning research, based on the R statistical programming language. The primary tool for developing machine learning models in this research was H2O, an open-source machine learning platform, known for its variety of predictive models and Java-based architecture. See Algorithm 1 for an overview of the Explainable AutoML setup.

---

**Algorithm 1** Pseudocode for Explainable AutoML Setup

**Input:** labeled training dataset $D1$, labeled test dataset $Dt$, maximum number of models $n$, metric used to sort the leaderboard (e.g. AUC) $L$

**Step 1:** Train Machine Learning Classifiers (GLM, DRF, XRT, GBM, DL)
**Step 2:** Repeat step 1 until the maximum number of models specified has been reached
**Step 3:** Use all pre-trained base classifiers to create stacking ensemble 1
**Step 4:** Use only the best classifier per category to create stacking ensemble 2
**Step 5:** Evaluate model performance based on specified metrics and create a leaderboard
**Step 6:** Run explainability methods on the leaderboard (e.g. Shapley Values, Variable Importance Heatmap)

**Output: a** set of visuals to provide guidance for human analysts to interpret model behavior, supporting informed decision-making and model validation.

---

The term credit decision denotes the process of classifying applicants for loans or credit cards into either a positive or negative category. The datasets used for the experiment have been extensively employed in prior research (Gunnarsson et al., 2021; Guo et al., 2019; Hamori et al., 2018; Schmitt & Cummins, 2023; Teng et al., 2013) and accurately reflect information typically accessible to retail banks, making them relevant for real-world scenarios. These datasets are publicly accessible, facilitating the reproducibility of the empirical analysis. Each dataset includes a target column indicating client defaults. See Table 1 for detailed information on each dataset and Table 2 for a description of the features.

Table 1. The empirical study is based on 2 data sets, each containing several features (predictors) including a target column containing the default information (response).

| Datasets | Observations | | | | | Description |
|---|---|---|---|---|---|---|
| | Total | y = 0 | y = 1 | Balanced* | Features | |
| Taiwan | 30,000 | 23,364 | 6,636 | 6,636/6,636 | 23 | Prediction whether a customer is going to default on their loan payment |
| Germany | 1,000 | 700 | 300 | 300/300 | 20 | Prediction whether a customer is going to default on their credit card payments |

*For the purpose of this study random under-sampling was used to bring the datasets in a balanced state

- **Dataset 1 – Taiwan:** The first dataset represents payment information from Taiwanese credit card clients. It was first used by Yeh and Lien (2009) and contains 30,000 observations where 6,636 are flagged as defaults. The dataset contains mainly historical payment information. Each observation (or feature set) contains 23 features including a binary response column for the default information of the credit cardholder.
- **Dataset 2 – Germany:** The second dataset represents detailed customer-level data from a German bank and contains 1,000 observations where 300 are flagged as defaults. Each observation contains 20 features across a diver's range of categories including a binary response column that indicates whether a particular client defaulted on their loan payments.

In the preprocessing phase of this study, several adjustments were made to optimize the dataset. To mitigate bias towards the majority class in this classification study, random under-sampling was employed to balance the dataset. This involved equalizing the class distribution between the 'good' and 'bad' categories. Additionally, categorical values in the data were converted to numerical form using one-hot encoding, a common technique for handling categorical data that converts labels into binary vectors. The response variable was also modified from a numerical to a binary format, suitable for classification. The dataset was divided into training and test sets with an 80:20 split. Finally, the evaluation metrics employed in this study to rank the models for the leaderboard is the area under the curve (AUC), which reflects a model's ability to distinguish between classes, with a score ranging from 1 (perfect) to 0.5 (random guessing).

**Table 2.** A detailed description of features contained in datasets 1 and 2.

| | Dataset 1 - Taiwan | | Dataset 2 - German |
|---|---|---|---|
| **Variable** | **Description** | **Variable** | **Description** |
| X1 | Amount of the given credit | X1 | Balance of checking account |
| X2 | Gender (1 = male; 2 = female) | X2 | Duration in months |
| X3 | Education* | X3 | Credit history |
| X4 | Marital status** | X4 | For what was the loan taken |
| X5 | Age (year) | X5 | Credit amount |
| X6 | Payment history September 2005 | X6 | Savings account plus bonds |
| X7 | Payment history August 2005 | X7 | Duration of current employment |
| X8 | Payment history July 2005 | X8 | Installment rate as % of income |
| X9 | Payment history June 2005 | X9 | Marital status and gender |
| X10 | Payment history May 2005 | X10 | Other debtors/guarantors |
| X11 | Payment history April 2005 | X11 | Present residence since |
| X12 | Amount of bill statement in Sep 2005 | X12 | Type of owned properties |
| X13 | Amount of bill statement in Aug 2005 | X13 | Age of applicant |
| X14 | Amount of bill statement in Jul 2005 | X14 | Housing (rent, own, free) |
| X15 | Amount of bill statement in Jun 2005 | X15 | Credits at other banks |
| X16 | Amount of bill statement in May 2005 | X16 | Existing credits at this bank |
| X17 | Amount of bill statement in Apr 2005 | X17 | Employment/Level of qualification |
| X18 | Amount paid September 2005 | X18 | The number of dependents |
| X19 | Amount paid August 2005 | X19 | Registered telephone or none |
| X20 | Amount paid July 2005 | X20 | Immigrant/foreign worker |
| X21 | Amount paid June 2005 | | |
| X22 | Amount paid May 2005 | | |
| X23 | Amount paid April 2005 | | |

* (1 = graduate school; 2 = university; 3 = high school; 4 = others)

** (1 = married; 2 = single; 3 = others)

## 3  Empirical Results

The results section focuses on four crucial figures that enhance explainability and transparency in the machine learning models developed using AutoML for credit decision-making in two datasets: Taiwan and German. These figures serve as instrumental tools for humans to comprehend the reasoning behind the models and identify the most influential features in the decision-making process.

The SHAP value plot for the Taiwan dataset's best-performing classifier (Figure 2), the Gradient Boosting Machine (GBM), provides an understanding of feature influence on model predictions. It visually represents how each feature contributes to shifting the prediction from a baseline, offering a transparent view of the factors driving credit decisions. This plot is crucial for interpretability, as it allows for the identification of key features that significantly impact the model's output. In this case, the most important features were the 'payment.history.sep2005', followed by the 'payment.history.aug2005', the 'payment.history.jun2005', and the 'credit.amount'.

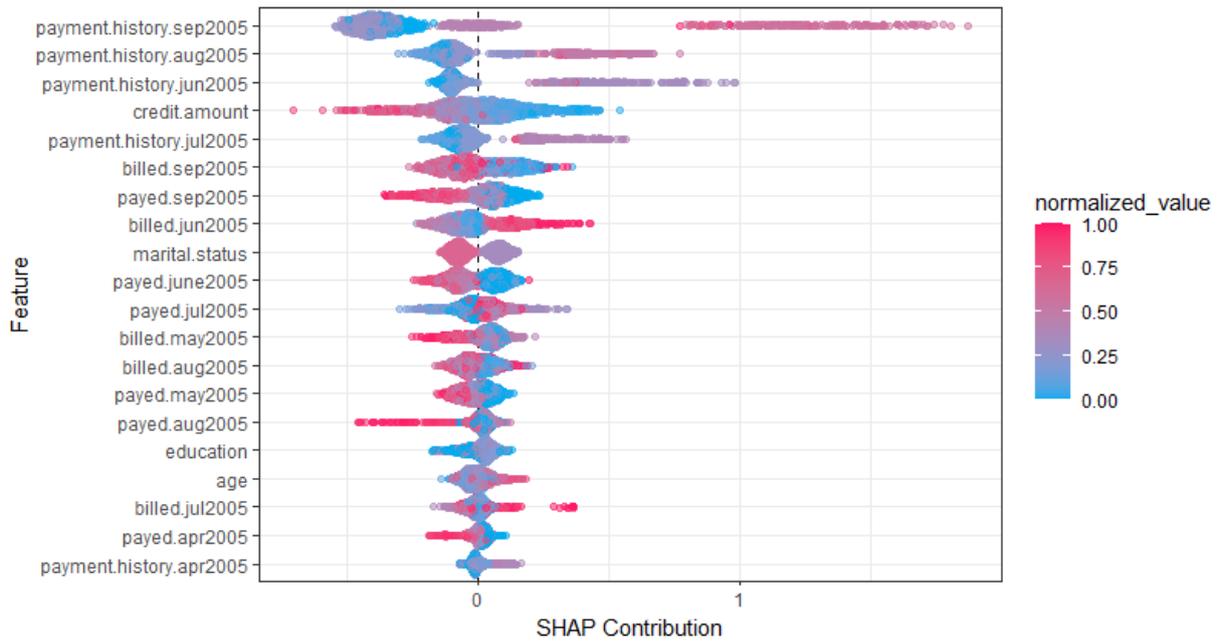

**Figure 2.** Feature attribution of the AutoML trained "leader" model (Taiwan)

Similar to the Taiwan model, the SHAP value plot for the German dataset's GBM classifier (Figure 3) shows the contribution of each feature in the predictive process. This enhances the transparency of the model, allowing stakeholders to see which attributes are most influential in credit decision-making in the German context. The most relevant features in this case were 'account.balance', 'credit.amount', 'previous.credit.payment.status', and 'credit.duration.months'.

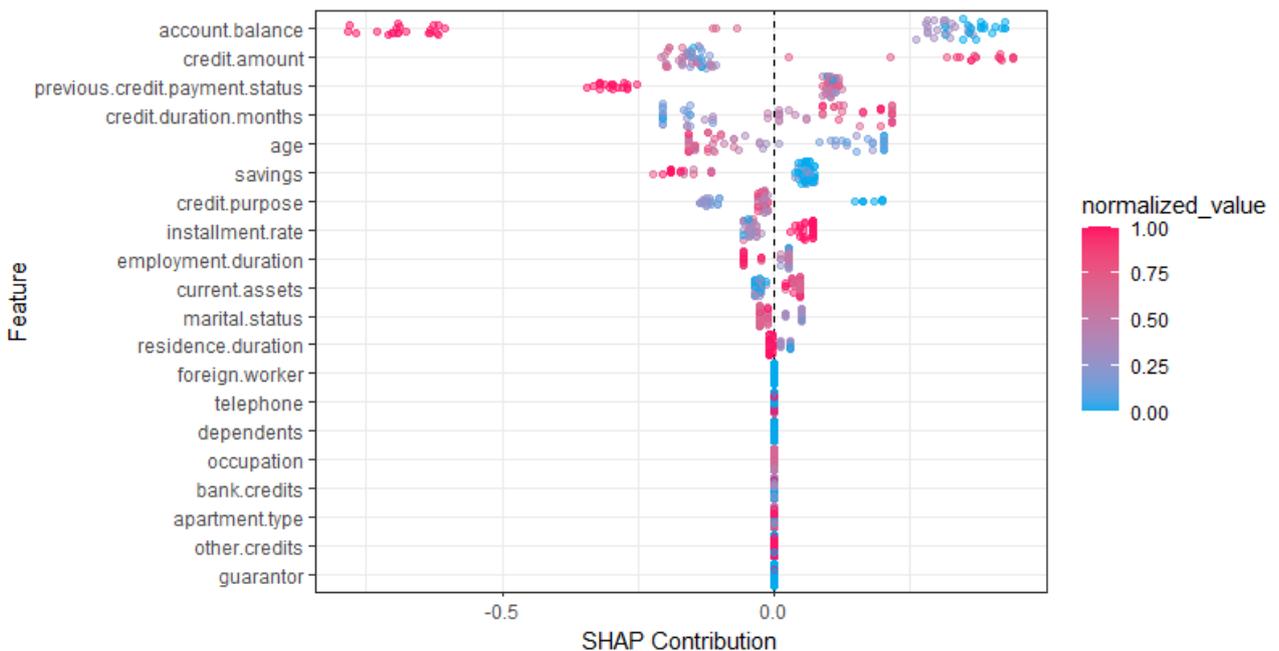

**Figure 3.** Feature attribution of the AutoML trained "leader" model (German)

The heatmap extends the scope of explainability by showing the variable importance across all classifiers generated by AutoML for the Taiwan dataset (Figure 3). It aggregates and compares the significance of each feature across various models, providing a comprehensive view of the most consistently influential factors. This insight is valuable for understanding the robustness of certain features in the predictive models, irrespective of the algorithm used. It can be observed that the 'payment.history.sep2005' is the most important feature across most models with a significant margin of error.

**Figure 4.** Variable importance heatmap across all AutoML-trained ML Models (Taiwan)

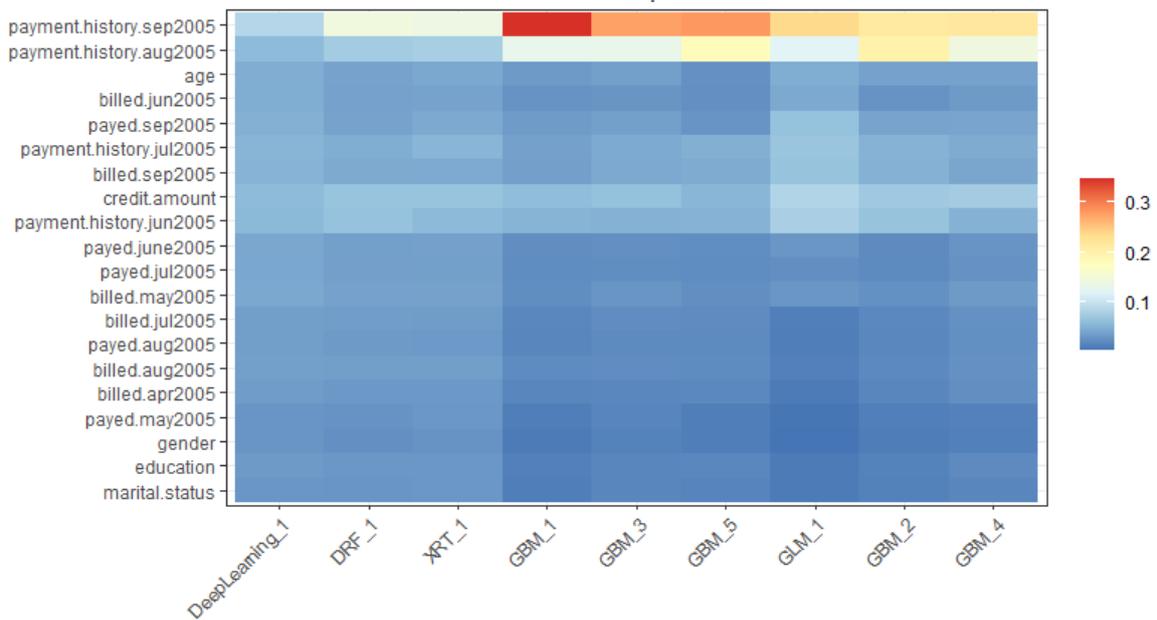

For the German dataset, this heatmap (Figure 4) mirrors the approach taken for the Taiwan dataset, highlighting the feature importance across all AutoML-trained models. It offers a broader perspective on the features that consistently play a pivotal role in different machine learning algorithms, reinforcing the understanding of what drives the models' decisions in the German credit scoring context. While 'account.balance' is still relevant across most models, the importance of this feature is not as significant as the 'payment.history.sep2005' in the Taiwan dataset.

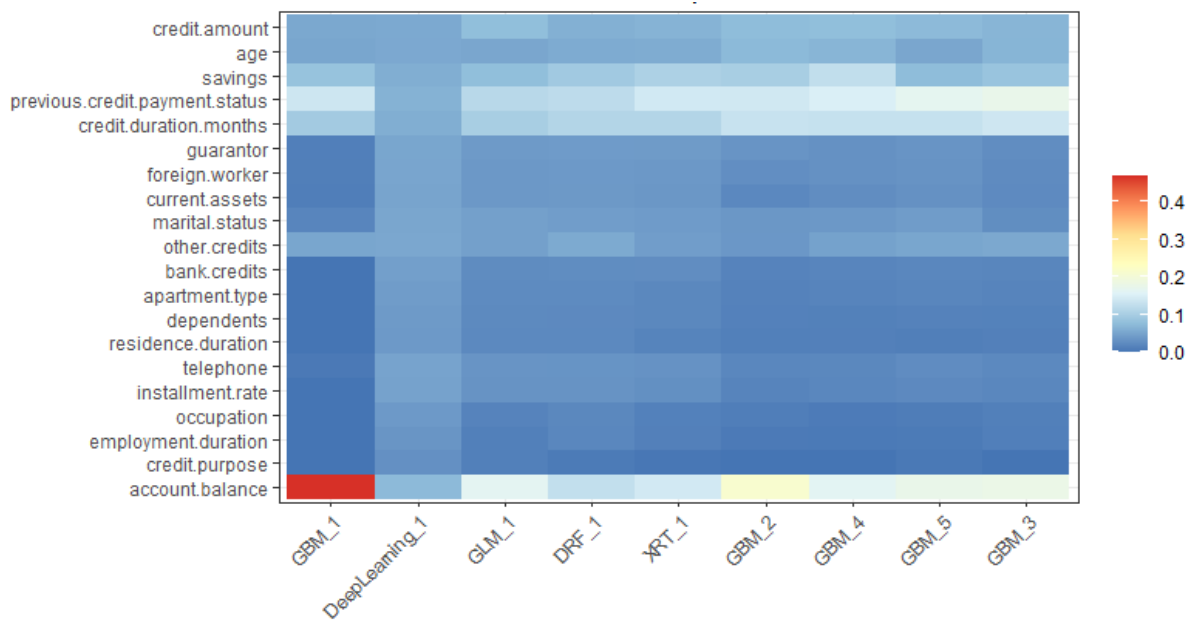

**Figure 5.** Variable importance across all AutoML trained ML Models (German)

# 4   Discussion

## 4.1   From Explainable AutoML to Human-AI Collaboration

Explainable AI (XAI), particularly methods like Shapley Values, plays a crucial role in the context of human-AI collaboration within AutoML frameworks. As AutoML automates the complex process of model building, the need for transparency becomes paramount to build trust and understanding among human users. Explainable AI bridges this gap by providing insights into how models make decisions, thereby fostering confidence and acceptance among stakeholders who may not be machine learning experts. For instance, Shapley Values can elucidate the contribution of each feature to a model's prediction, enabling users to comprehend and validate the automated decisions. This transparency is essential in all sensitive areas such as healthcare, finance, and legal systems, where understanding the rationale behind AI-driven decisions is as important as the decisions themselves (Kleinberg et al., 2017; Schmitt & Cummins, 2023; Topol, 2019). Therefore, integrating XAI into AutoML facilitates more effective human-AI collaboration, ensuring that automated models are not only efficient but also transparent and accountable.

## 4.2   Implications for Practice

This research highlights the critical role of Automated Machine Learning (AutoML) and Explainable AI, with a specific focus on Shapley Values, in shaping credit decisions within the digital economy. In domains like financial services, it is imperative to look beyond mere accuracy to include model transparency and resource efficiency. While Explainable AutoML can assist in enhancing human-AI collaboration and democratizing AI/ML for non-experts, XAI is particularly important in the context of credit decisions and credit risk management for several key reasons:

- **Regulatory Compliance:** Financial institutions are often subject to stringent regulatory requirements that mandate transparency in their decision-making processes. XAI can potentially enable these institutions to provide explanations of credit decisions made by AI systems, which is essential for complying with regulations like the Equal Credit Opportunity Act (ECOA) and General Data Protection Regulation (GDPR), which require decisions to be explainable and non-discriminatory.
- **Risk Assessment Accuracy:** In credit risk modeling, understanding the factors that influence a model's decision is crucial for accuracy and reliability. XAI techniques like Shapley Values can help in identifying which variables (e.g., income, credit history) are most influential in assessing creditworthiness. This insight can lead to more accurate and fair credit risk assessments.
- **Building Trust with Customers:** Transparency in credit decision processes helps in building trust with customers. When individuals are denied credit or given less favorable terms, they are more likely to accept the decision if they understand the reasons behind it. This transparency can enhance customer relationships and trust in the financial institution.
- **Ethical Considerations and Bias Mitigation:** XAI aids in uncovering and addressing potential biases in credit decision models. By understanding how and why certain decisions are made, institutions can identify and correct biases against certain groups, ensuring fair and ethical lending practices.
- **Continuous Improvement:** XAI provides feedback that is essential for the continuous improvement of credit risk models. By understanding model behaviors, data scientists and risk managers can refine and adjust models to better capture the complexities of credit risk.

## 4.3 Future Research

Future research should thus not only focus on enhancing accuracy with sophisticated data and methods but also on fostering AI systems that are comprehensible, collaborative, and sustainable, ensuring they are aligned with the ethical and practical requirements of the evolving digital economy. See Table 3.

**Table 3.** Future Research Directions in AI for Credit Decision-Making

| Research Theme | Description |
|---|---|
| Human-AI Collaboration | Research should explore enhanced human-in-the-loop systems (Capel & Brereton, 2023; Wu et al., 2022) to find the most effective ways for human experts to interact with AI in credit |

| | decision-making processes. This aims to optimize the synergy between human intuition and AI's analytical capabilities. |
|---|---|
| Robustness and Security | Future studies need to ensure the robustness of AI systems against potential manipulations (Schmitt & Koutroumpis, 2023) and thoroughly understand the security implications in credit scoring applications. |
| Generative AI and Synthetic Data Generation | Utilizing generative AI to create synthetic financial profiles can expand training datasets for credit scoring models while preserving customer privacy. In addition, generative AI can be used for novel use cases. |
| Sustainability of AI/ML Models | The environmental impact of AI/ML models, especially the carbon emissions from training complex models like deep learning, is a growing concern (Dhar, 2020). Future research should focus on integrating renewable energy sources, developing energy-efficient algorithms, and innovations in hardware and data center designs. This is crucial for aligning AI advancements with environmental responsibilities and initiatives like the European Green Deal. |
| From Explainability to Interpretability | While XAI has made strides in enhancing AI transparency, achieving full interpretability, particularly in complex models like Deep Learning, remains a challenge. Research must continue to integrate XAI improvements into credit scoring models, striving for complete interpretability and auditability to meet regulatory standards and gain broader acceptance. |

# 5 Conclusion

Fair credit decisions are an essential building block of the economy. This paper presented a detailed examination of the use of Explainable Automated Machine Learning (AutoML) in credit decision-making. The integration of AutoML with explainable AI (XAI) methods, particularly using SHAP value plots and variable importance heatmaps, has illustrated the path towards more transparent, and human-centric decision-making in credit scoring. By allowing a glimpse into the inner workings of machine learning models, these tools help to bridge the gap between AI/ML Blackbox algorithms, and the understanding required for effective Human-AI collaboration. The results underscore the potential of AutoML to not only automate the process of model creation but also to enhance the human role in supervising, understanding, and ethically guiding AI decisions.